\documentclass{article}

\usepackage{arxiv}

\usepackage[utf8]{inputenc} % allow utf-8 input
\usepackage[T1]{fontenc}    % use 8-bit T1 fonts
\usepackage{hyperref}       % hyperlinks
\usepackage{url}            % simple URL typesetting
\usepackage{booktabs}       % professional-quality tables
\usepackage{amsfonts}       % blackboard math symbols
\usepackage{nicefrac}       % compact symbols for 1/2, etc.
\usepackage{microtype}      % microtypography
\usepackage{lipsum}
\usepackage{graphicx}
\usepackage{epstopdf}
 \title{Complex network analysis of Indian  railway zones}

\author{
  Nikhil Kumar Rajput\\ %\thanks{}
  Department of Computer Science\\
  Ramanujan College, University of Delhi\\
  New Delhi India \\
  \texttt{n.rajput@ramanujan.du.ac.in} \\
 \And
 Piyush Badola \\
  Department of Computer Science\\
  Ramanujan College, University of Delhi\\
  New Delhi India \\
  \texttt{piyush.badola@gmail.com } \\
   \And
 Harshit Arora \\
  Department of Computer Science\\
  Ramanujan College, University of Delhi\\
  New Delhi India \\
  \texttt{coolharshit149@gmail.com } \\
\And
 Bhavya Ahuja Grover \\
  Department of Computer Science\\
  Ramanujan College, University of Delhi\\
  New Delhi India \\
  \texttt{b.ahuja@ramanujan.du.ac.in } \\
}

\begin{document}
\maketitle

\begin{abstract}
Indian Railway Network has been analyzed on the basis of number of trains directly linking two railway zones. The network has been displayed as a weighted graph where the weights denote the number of trains between the zones. It may be pointed out that each zone is a complex network in itself and may depict different characteristic features. The zonal network therefore can be considered as a network of complex networks. In this paper, self links, in-degree and out-degree of each zone have been computed which provides information about the inter and intra zonal connectivity. Degree-passenger correlation which gives an idea about number of trains and passengers originating from a particular zone which might play a role in policy making decisions has also been studied. Some other complex network parameters like betweenness, clustering coefficient and cliques have been obtained to get more insight about the complex Indian zonal network.
\end{abstract}

% keywords can be removed

\section{Introduction}
Indian railways is one of the largest transport networks in the world. It has about 67,368 kms of tracks and  about 8116 million passengers originating with an average distance travelled of 141.6 kms per person in 2016-17(\url{indianrail.gov.in}). The railway network is structured into multi-tier system. The zones are at top of the hierarchy which are divided into divisions\cite{abraham2008performance}. The division controls the railway stations which is the last entity in the hierarchy. At present there are 18 zones with 73 divisions and 7349 stations(\url{indianrail.gov.in}).\\
Transportation system in a complex network framework has been studied in airlines\cite{bagler2008analysis}, railways\cite{li2007empirical,sen2003small}, subways\cite{angeloudis2006large} and streetways\cite{porta2006network}. Scale free properties have also been observed in railway networks\cite{li2007empirical,sen2003small}. Complex network based fault spreading model was investigated in \cite{zhou2015railway} to analyse the impact of failures in railway networks. A multilayer complex weighted network model for understanding spatio-temporal traffic flow patterns in Chinese subway network was given in \cite{feng2017weighted}. Characterization of train flow and passenger flow network was also carried out in their work. A study was done on the rail and bus transportation services as a complex weighted network in Singapore to get insights about its typological and dynamic pattern \cite{soh2010weighted}.\\
This work aims at presenting an overview of the zonal network that builds the Indian Railways. The data has been collected from (\url{indianrail.gov.in}) which comprises the number of trains originating from one railway zone and arriving at the other zones or the same zone. A complex weighted graph has been generated using the data. The network graph has been analyzed and some significant deductions have been made which are discussed in the subsequent sections. The number of passengers traveling from one zone to another has also been obtained from (\url{data.gov.in}) which has been used to find the correlation with the number of trains in a particular zone.
\begin{figure}[h]
  \centering
  \includegraphics[width=18cm,height=15cm]{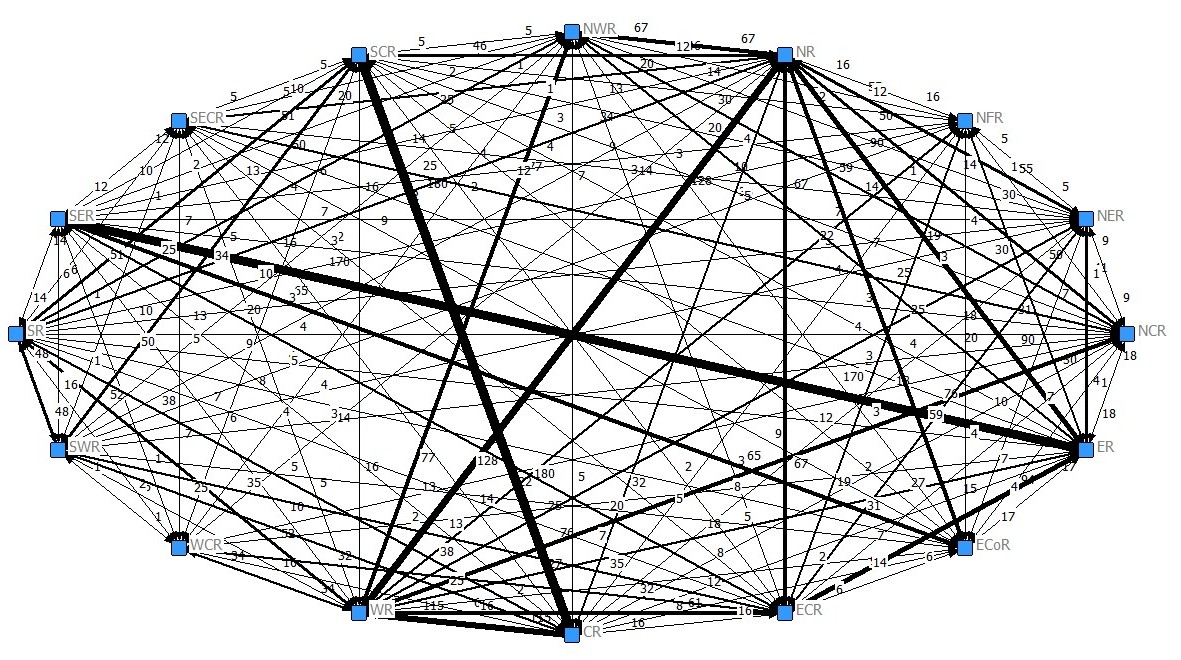}\\
  \caption{Complex India Zonal Railway Network}\label{zone}
\end{figure}
\section{Complex Indian Railway Zonal Network}
A number of trains ply from one railway zone to another as well as connect stations within a particular zone. The inter-zone connectivity has been depicted in the figure \ref{zone} as a weighted graph. The weights denote the number of trains between the zones. The width of the lines also corresponds to their weights.
\begin{table}[htb]
\caption{Structural properties of Indian Zonal Network }\label{IRNZonestable}
\begin{tabular}{|p{2cm}|p{1.5cm}|p{1.5cm}|p{1.5cm}|p{1.5cm}|p{1.5cm}|p{1.5cm}|p{1.5cm}|}
\hline
Zones	&	Out	&	In	&	Clustering 	&	Betwe-	&	Self	&	Total 	&Passeng-\\
	&	degree	&	degree	&	 Coefficient	&	enness	&	 Links	&	Links &ers 2014 	\\
&		&		&	 	&		&	 	&	 &-15(in	\\
&		&		&	 	&		&	 	&	 &millions)	\\
\hline
CR	&	509	&	528	&	21.076	&	0.812	&	3694	&	4203	&	1716\\
ER	&	535	&	531	&	21.863	&	0.669	&	3450	&	3985	&	1156\\
WR	&	652	&	652	&	19.805	&	0.812	&	2840	&	3492	&	1625\\
SR	&	310	&	313	&	23.048	&	0.812	&	2886	&	3196	&	806	\\
NR	&	706	&	704	&	19.3	&	0.812	&	1308	&	2014	&	637	\\
SCR	&	392	&	393	&	22.276	&	0.812	&	1160	&	1552	&	357	\\
SER	&	360	&	360	&	22.586	&	0.812	&	614	&	974	&	256	\\
NER	&	236	&	234	&	25.94	&	0.297	&	584	&	820	&	165	\\
ECR	&	402	&	402	&	22.186	&	0.812	&	360	&	762	&	254	\\
NWR	&	279	&	279	&	23.367	&	0.812	&	386	&	665	&	153	\\
NFR	&	137	&	134	&	29.987	&	0.143	&	478	&	615	&	87	\\
SWR	&	209	&	209	&	21.863	&	0.669	&	362	&	571	&	191	\\
NCR	&	256	&	254	&	23.586	&	0.812	&	234	&	490	&	171	\\
SECR	&	129	&	127	&	27.121	&	0.476	&	350	&	479	&	125	\\
ECoR	&	215	&	209	&	26.192	&	0.297	&	216	&	431	&	90	\\
WCR	&	136	&	136	&	30.742	&	0.143	&	96	&	232	&	139	\\
\hline
\end{tabular}
%%use \tablenotes{footnote} to get the table foot note
%\tablenotes{Sample table footnote}
\end{table}

\subsection{Self Links, In and Out Degrees}
The number of links varies considerably in inter-zone as well as intra-zonal networks. There are three types of links viz. outgoing links, incoming links and self links. Here, the self links denote the trains between stations of a particular zone.\\
The total number of links comprise of the sum of outgoing links, incoming links and twice the number of self links. The highest out and in links have been observed for Northern Railways zone with values 706 and 704 respectively whereas the maximum value of self links is for CR zone with a value of 3694(See table\ref{IRNZonestable}).
\begin{figure}[ht]
  \centering
  \includegraphics[width=\textwidth,height=10cm]{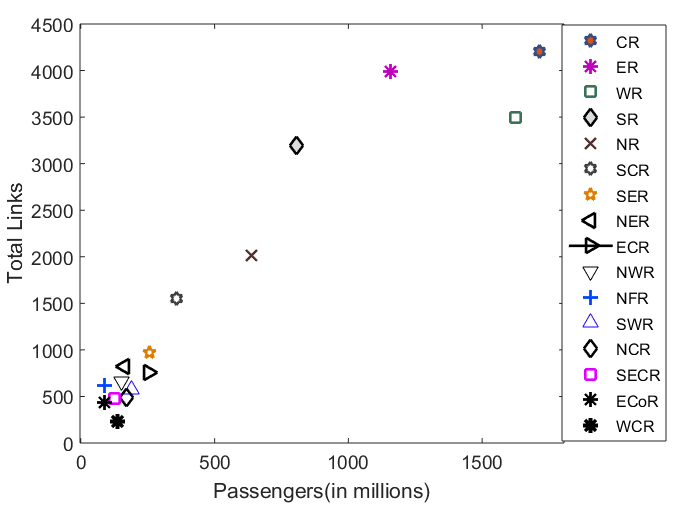}\\
  \caption{Degree-Passengers Correlation}\label{degree-passenger}
\end{figure}

\subsection{Betweenness, Clustering Coefficient and Cliques}
Betweeness centrality(BC) and clustering coefficient are important quantities which characterize a complex network. BC denotes how influential a node is in a network \cite{barthelemy2004betweenness}. On the other hand, clustering coefficient\cite{soffer2005network} provides the clustering tendency of a node in a network. Both these quantities have been given in table\ref{IRNZonestable}) for each zone. The highest BC has been noted to be 0.812 which is for CR,WR,SR,NR,SCR,SER,ECR,NWR and NCR and lowest being 0.143 for NFR and WCR. The clustering coefficient is found to be maximum for WCR with 30.742 and minimum for NR with a value of 19.3.\\
Cliques denote the direct connectivity between every node of subgraphs of a graph network. There are six cliques observed in Zonal network of Indian railways. It can be observed that NR,NWR,SR,CR,SER,SCR,NCR,ECR, and WR zones appeared in every clique.
\subsection{Degree-Passengers and Degree-Degree Correlations}
It is imperative to see the correlation between the total number of trains and the passengers originating from that zone. There is an obvious positive correlation with highest pair value for CR as 4203 trains and 1716 million passengers in the year 2014-15. Figure \ref{degree-passenger} shows the degree-passenger correlation for all the zones.
\begin{figure}[ht]
  \centering
  \includegraphics[width=\textwidth]{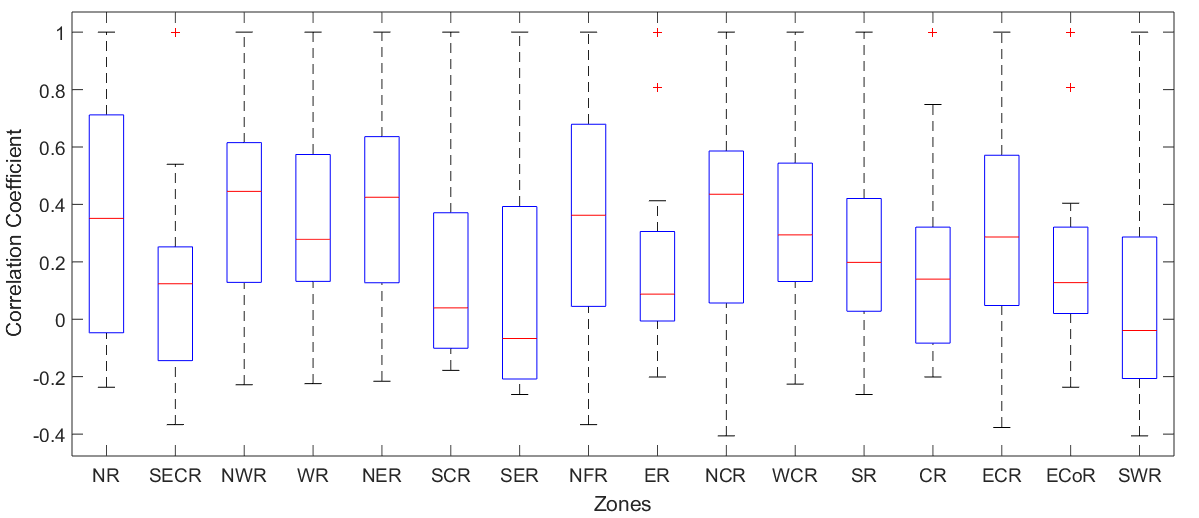}\\
  \caption{Box plot of degree-degree correlation coefficients.}\label{corr}
\end{figure}
The degree correlation between the zones is shown in the figure \label{corr}.
 The box plot gives a brief idea about the quartiles, the range and the mean values of degree-degree correlations of different zones. It can be also visualized that the correlation matrix shows positive values with a significant number of negative values which correspond to the zones which are farther from each other (see table \ref{degreecorrelation}). The highest range of correlation values is for NR and highest mean value is for NWR.

%\section{References}
%\bibliographystyle{elsarticle-harv}
%\bibliography{irnref}

%\newpage

\begin{table*}[htb]
  \centering
  \caption{Degree-Degree Correlation}\label{degreecorrelation}
\begin{tabular}{|p{0.7cm}|p{0.68cm}|p{0.68cm}|p{0.68cm}|p{0.68cm}|p{0.68cm}|p{0.68cm}|p{0.68cm}|p{0.68cm}|p{0.68cm}|p{0.68cm}|p{0.68cm}|p{0.68cm}|p{0.68cm}|p{0.68cm}|p{0.68cm}|p{0.68cm}|}
\hline
 $\setminus$&NR&SECR&NWR&WR&NER&SCR&SER&NFR&ER&NCR&WCR&SR&CR&ECR&ECoR&SWR\\
 \hline
 NR&1&-0.06&0.73&0.28&0.69&-0.16&0.37&0.76&-0.03&0.72&0.57&0.18&0.32&0.51&-0.23&-0.15\\
 \hline
 SECR&-0.06&1&0.29&0.54&0.19&0.19&-0.24&-0.36&0.20&0.21&0.05&-0.15&-0.13&-0.10&0.38&-0.19\\
 NWR&0.73&0.29&1&0.65&0.57&-0.03&-0.06&0.51&0.06&0.90&0.51&0.38&0.28&0.55&0.18&-0.22\\
 WR&0.28&0.54&0.65&1&0.56&0.52&-0.22&0.27&0.10&0.58&0.59&0.21&-0.09&0.15&0.16&0.09\\
 NER&0.69&0.19&0.57&0.56&1&0.05&0.40&0.76&0.41&0.56&0.43&0.02&-0.07&0.76&0.40&-0.21\\
 SCR&-0.16&0.19&-0.03&0.52&0.05&1&-0.12&-0.11&-0.17&-0.08&0.14&0.53&0.31&-0.06&0.02&0.42\\
 SER&0.37&-0.24&-0.06&-0.22&0.40&-0.12&1&0.59&0.04&-0.06&-0.22&-0.26&-0.14&0.59&0.16&-0.19\\
 NFR&0.76&-0.36&0.51&0.27&0.76&-0.11&0.59&1&0.30&0.49&0.42&0.07&0.01&0.81&0.08&-0.09\\
ER&-0.03&0.20&0.06&0.10&0.41&-0.17&0.04&0.30&1&0.02&0.12&-0.18&-0.20&0.30&0.80&0.01\\
NCR&0.72&0.21&0.90&0.58&0.56&-0.08&-0.06&0.49&0.02&1&0.58&0.37&0.26&0.49&0.08&-0.40\\
WCR&0.57&0.05&0.51&0.59&0.43&0.14&-0.22&0.42&0.12&0.58&1&0.28&0.30&0.26&-0.15&0.14\\
SR&0.18&-0.15&0.38&0.21&0.02&0.53&-0.26&0.07&-0.18&0.37&0.28&1&0.74&0.09&0.02&0.46\\
CR&0.32&-0.13&0.28&-0.09&-0.07&0.31&-0.14&0.01&-0.20&0.26&0.30&0.74&1&0.001&-0.05&0.50\\
ECR&0.51&-0.10&0.55&0.15&0.76&-0.06&0.59&0.81&0.30&0.49&0.26&0.09&0.001&1&0.26&-0.37\\
ECoR&-0.23&0.38&0.18&0.16&0.40&0.02&0.16&0.08&0.80&0.08&-0.15&0.02&-0.05&0.26&1&0.01\\
SWR&-0.15&-0.19&-0.22&0.09&-0.21&0.42&-0.19&-0.09&0.01&-0.40&0.14&0.46&0.50&-0.37&0.01&1 \\
\hline
  %\caption{}\label{}
  \end{tabular}
\end{table*}

\section{Conclusion}
A weighted complex network of Indian railway zones was considered for the study. The zonal network comprises of links among the stations in that zone as well as trains from one zone to another. The graphical visualization of the network has been presented with weights denoting the number of trains. Characterizing features for gaining more insight about the network have also been presented. Betweenness centrality, clustering coefficient, in degree, out degree, self links have been calculate that give a brief idea about the network. Degree-Passenger and Degree-Degree correlations provide detail about correspondence between number of trains with passengers and the degree of connectivity between zones respectively.

%Cliques\\
%
%1.\quad  NR - SECR - NWR - WR - NER - SCR - SER - ER - NCR - SR - CR - ECR - ECoR \\
%    2. \quad NR - SECR - NWR - WR - NER - SCR - SER - NCR - SR - CR - ECR - ECoR - SWR  \\
%    3. \quad NR - SECR - NWR - WR - SCR - SER - ER - NCR - WCR - SR - CR - ECR  \\
%    4. \quad NR - SECR - NWR - WR - SCR - SER - NCR - WCR - SR - CR - ECR - SWR  \\
%    5. \quad NR - NWR - WR - NER - SCR - SER - NFR - ER - NCR - SR - CR - ECR - ECoR  \\
%    6. \quad NR - NWR - WR - NER - SCR - SER - NFR - NCR - SR - CR - ECR - ECoR - SWR \\
\clearpage
\bibliographystyle{unsrt}
\bibliography{irn}  %%% Remove comment to use the external .bib file (using bibtex).


\end{document}